\documentclass[runningheads]{llncs}
\usepackage[T1]{fontenc}
\usepackage{graphicx,verbatim}
\usepackage{hyperref}
\usepackage{bbding}
\usepackage{amsmath}
\usepackage{amsfonts}
\usepackage{array}
\usepackage{xcolor}
\usepackage{multirow}
\usepackage{booktabs}
\usepackage{cite}

\usepackage{amssymb}

\usepackage{diagbox}
\usepackage[graphicx]{realboxes}
 
\begin{document}
\title{GL-LCM: Global-Local Latent Consistency\\Models for Fast High-Resolution Bone\\Suppression in Chest X-Ray Images}
\titlerunning{Global-Local Latent Consistency Models for Bone Suppression}
\begin{comment}  %% Removed for anonymized MICCAI 2025 submission
\author{First Author\inst{1}\orcidID{0000-1111-2222-3333} \and
Second Author\inst{2,3}\orcidID{1111-2222-3333-4444} \and
Third Author\inst{3}\orcidID{2222--3333-4444-5555}}
%
\authorrunning{F. Author et al.}
\institute{Princeton University, Princeton NJ 08544, USA \and
Springer Heidelberg, Tiergartenstr. 17, 69121 Heidelberg, Germany
\email{lncs@springer.com}\\
\url{http://www.springer.com/gp/computer-science/lncs} \and
ABC Institute, Rupert-Karls-University Heidelberg, Heidelberg, Germany\\
\email{\{abc,lncs\}@uni-heidelberg.de}}

\end{comment}

\author{Yifei Sun\inst{1} \and
Zhanghao Chen\inst{1} \and
Hao Zheng\inst{1} \and
Yuqing Lu\inst{2} \and
Lixin Duan\inst{3} \and
\\
Fenglei Fan\inst{4} \and
Ahmed Elazab\inst{5} \and
Xiang Wan\inst{6} \and
Changmiao Wang\inst{6,}\textsuperscript{\Envelope} \and
\\
Ruiquan Ge\inst{1,}\textsuperscript{\Envelope}
}
% index{Sun, Yifei}
% index{Chen, Zhanghao}
% index{Zheng, Hao}
% index{Lu, Yuqing}
% index{Duan, Lixin}
% index{Fan, Fenglei}
% index{Elazab, Ahmed}
% index{Wan, Xiang}
% index{Wang, Changmiao}
% index{Ge, Ruiquan}
%
\authorrunning{Y. Sun et al.}
\institute{Hangzhou Dianzi University, Hangzhou, China\\ 
\email{gespring@hdu.edu.cn}
\and The Chinese University of Hong Kong, Hong Kong, China \and University of Electronic Science and Technology of China, Shenzhen, China \and City University of Hong Kong, Hong Kong, China \and Shenzhen University, Shenzhen, China \and Shenzhen Research Institute of Big Data, Shenzhen, China\\
\email{cmwangalbert@gmail.com}
}

\maketitle              % typeset the header of the contribution
\begin{abstract}
Chest X-Ray (CXR) imaging for pulmonary diagnosis raises significant challenges, primarily because bone structures can obscure critical details necessary for accurate diagnosis. Recent advances in deep learning, particularly with diffusion models, offer significant promise for effectively minimizing the visibility of bone structures in CXR images, thereby improving clarity and diagnostic accuracy. Nevertheless, existing diffusion-based methods for bone suppression in CXR imaging struggle to balance the complete suppression of bones with preserving local texture details. Additionally, their high computational demand and extended processing time hinder their practical use in clinical settings. To address these limitations, we introduce a Global-Local Latent Consistency Model (GL-LCM) architecture. This model combines lung segmentation, dual-path sampling, and global-local fusion, enabling fast high-resolution bone suppression in CXR images. To tackle potential boundary artifacts and detail blurring in local-path sampling, we further propose Local-Enhanced Guidance, which addresses these issues without additional training. Comprehensive experiments on a self-collected dataset SZCH-X-Rays, and the public dataset JSRT, reveal that our GL-LCM delivers superior bone suppression and remarkable computational efficiency, significantly outperforming several competitive methods. Our code is available at \href{https://github.com/diaoquesang/GL-LCM}{https://github.com/diaoquesang/GL-LCM}.

% Authors must provide keywords and are not allowed to remove this Keyword section.

\end{abstract}

\section{Introduction}
Lung disease remains a major global health challenge, contributing to high rates of morbidity and mortality \cite{pham2021cnn,zhen2025epidemiology}. Among available diagnostic tools, Chest X-Ray (CXR) is widely regarded as the primary imaging modality for evaluating pulmonary conditions such as inflammation, tuberculosis, and lung masses. This preference is due to its accessibility, affordability, and low radiation exposure. However, interpreting CXR images poses notable difficulties. Even experienced radiologists may overlook lesions because superimposed bone structures can obscure critical diagnostic information \cite{liu2023bone}. In light of these challenges, the development of bone suppression techniques in CXR imaging offers great promise for enhancing the diagnostic capabilities of radiologists \cite{li2020high} and improving the performance of computer-aided lung lesion detection systems \cite{wang2025rib}. Currently, Dual-Energy Subtraction (DES) imaging is the most commonly employed bone suppression technique in clinical practice. By utilizing two X-ray exposures at different energy levels, DES effectively reduces the visual clutter caused by overlapping bones. However, DES imaging requires specialized equipment and increases radiation exposure, making it less accessible and impractical in resource-limited settings. 

Conventional statistical-based methods for bone suppression in CXR imaging \cite{simko2009elimination, juhasz2010segmentation} offer an alternative to specialized DES equipment. However, these methods require precise segmentation and boundary annotation, hampering their practical applicability. In contrast, recent advancements in deep learning have enabled models to automatically learn complex image features, providing more accurate bone suppression. This has led to the development of advanced techniques for bone suppression. For instance, Gusarev \textit{et al.} \cite{gusarev2017deep} utilized an Auto-Encoder (AE) to generate soft tissue images from CXR images. Similarly, Zhou \textit{et al.} \cite{zhou2019generation} proposed a Multi-scale Conditional Adversarial Network (MCA-Net) drawing inspiration from Generative Adversarial Networks (GANs) \cite{goodfellow2014generative}. Additionally, Wang \textit{et al.} \cite{wang2025rib} introduced an improved U-Net with self-attention. Despite these efforts, performance limitations persist, limiting their applications in real-world clinical settings. 
% While these deep learning methods show promise, further refinement is necessary for effectiveness and reliability in real-world clinical applications.

Recently, diffusion models have demonstrated strong capabilities in various generation tasks \cite{wolleb2024binary,fu2024geowizard, ye2025tfg}. Although some studies have successfully used diffusion models for bone suppression in CXR imaging \cite{chen2024bs, sun2024bs}, applying these models to perform fast high-resolution bone suppression still encounters challenges given the crucial requirement in clinical settings \cite{weber2023cascaded}. Particularly, these methods, despite being end-to-end, fail to balance global suppression of bone structures with local detail retention, primarily due to the difficulty of processing global bone residuals and local detail features through a single network. Additionally, existing diffusion models for bone suppression are computationally demanding, requiring significant processing time, which is impractical for clinical applications.

To overcome these challenges, we propose the Global-Local Latent Consistency Model (GL-LCM), a novel framework for fast high-resolution bone suppression in CXR images based on Latent Consistency Models (LCMs) \cite{luo2023latent}. Our key contributions are summarized as follows: 
\underline{First}, the GL-LCM architecture facilitates effective bone suppression while retaining texture details. This is achieved through the design of dual-path sampling in the latent space combined with global-local fusion in the pixel space. 
\underline{Second}, GL-LCM significantly enhances inference efficiency, which requires only approximately 10\% of the inference time of current diffusion-based methods, making it more suitable for clinical applications.
\underline{Third}, we introduce Local-Enhanced Guidance (LEG) to mitigate potential boundary artifacts and detail blurring issues in local-path sampling without additional training.
\underline{Fourth}, extensive experiments on both the self-collected dataset SZCH-X-Rays and the public dataset JSRT demonstrate exceptional performance and efficiency of our GL-LCM.

\section{Proposed Method}

\begin{figure}[!tbp]
\centering
\includegraphics[width=\textwidth]{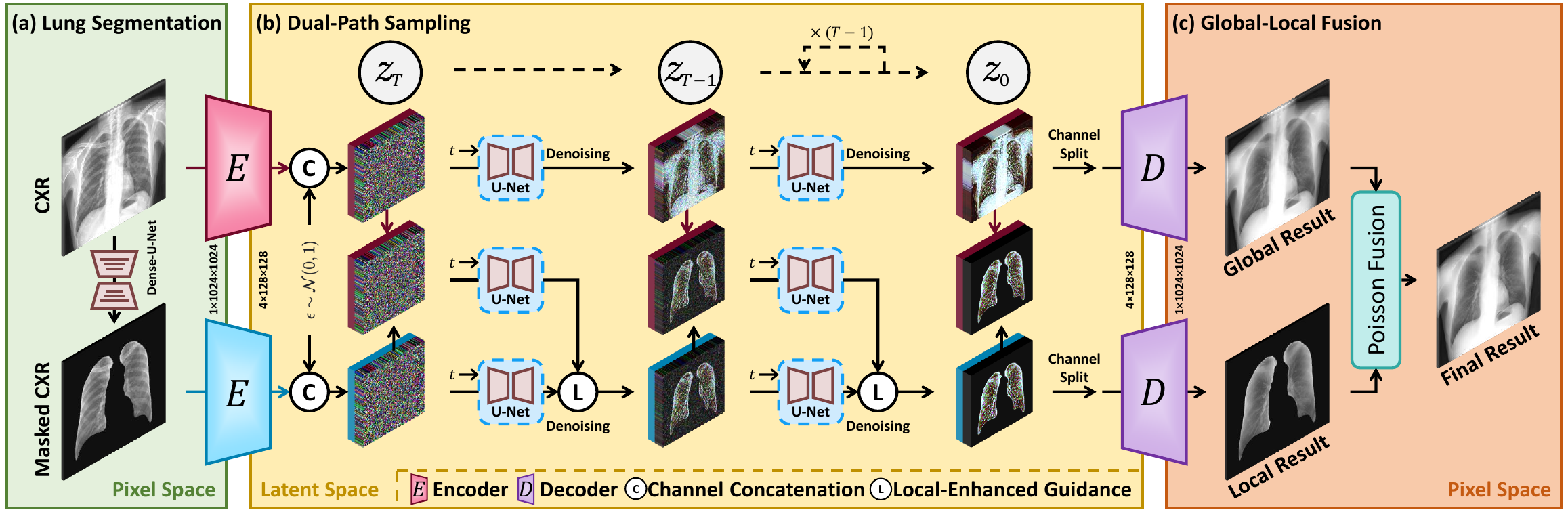}
\caption{Overview of GL-LCM framework. (a) Lung segmentation in the pixel space, (b) Dual-path sampling in the latent space, and (c) Global-local fusion in the pixel space.
}
\label{frame}
\end{figure}

The proposed GL-LCM framework, illustrated in Fig. \ref{frame}, operates in three key  stages. Initially, lung regions are segmented from CXR images in the pixel space using a pre-trained Dense-U-Net \cite{yahyatabar2020dense}. Next, using LEG, dual-path sampling of conditional LCMs is employed to produce both global and local soft tissue data. Finally, these global and local outputs are decoded back into the pixel space and integrated through Poisson Fusion to yield the final soft tissue image.

\subsection{Conditional Latent Consistency Models}
The LCMs \cite{luo2023latent} represent a new class of generative models that efficiently produce high-resolution images while preserving quality. Unlike traditional consistency models operating in the pixel space \cite{song2023consistency}, LCMs reduce computational demands by transforming data from the pixel space \(X\) to a latent space \(Z\). Informed by the principles of LCMs, our GL-LCM model incorporates both a forward and a reverse process within the latent space.

The forward process of our GL-LCM can be defined as:
\begin{equation}
z_t=\sqrt{\bar{\alpha}_t} z_0+\sqrt{1-\bar{\alpha}_t} \epsilon, \quad \epsilon \sim \mathcal{N}(\mathbf{0}, \mathbf{I}),
\label{eq_noise}
\end{equation}
where $\epsilon$ is a noise component sampled from a Gaussian distribution, and $\bar{\alpha}_t:=\prod_{s=0}^t \alpha_s$. Here, $\alpha_t=1-\beta_t$ is a differentiable function of timestep $t$ determined by the denoising LCM sampler. The training loss of GL-LCM is expressed as:
\begin{equation}
\mathcal{L}\left(\epsilon_\theta\right)=\sum_{t=1}^T \mathbb{E}_{z_0, \epsilon}\left[\left\|\epsilon_\theta\left(z_t, t, \widetilde{z}\right)-\epsilon\right\|_2^2\right],
\label{eq_loss}
\end{equation}
where $\epsilon_\theta$ represents the predicted noise at timestep $t$ by the noise estimator network with parameters $\theta$, and $\widetilde{z}$ is a given image condition.

During the reverse process of GL-LCM, starting from random noise $z_T \sim \mathcal{N}(\mathbf{0}, \mathbf{I})$, the final result $z_0$ is predicted through a multi-step denoising process:
\begin{equation}
z_{t-1}= 
\begin{cases}
\sqrt{\alpha_{t-1}}\left(c_{\text {out}}(t) \frac{z_t-\frac{1-\alpha_t}{\sqrt{1-\bar{\alpha}_t}} \epsilon_\theta\left(z_t, t, \tilde{z}\right)}{\sqrt{\alpha_t}}+c_{\text {skip}}(t) z_t\right)+\frac{1-\alpha_{t-1}}{\sqrt{1-\bar{\alpha}_{t-1}}} \epsilon, & \!\!1<t \leqslant T \\
c_{\text {out}}(t) \frac{z_t-\frac{1-\alpha_t}{\sqrt{1-\bar{\alpha}_t}} \epsilon_\theta\left(z_t, t, \tilde{z}\right)}{\sqrt{\alpha_t}}+c_{\text {skip}}(t) z_t, & \!\!t = 1,
\end{cases}
\label{eq_inf}
\end{equation}
where $T$ is the total sampling timesteps, and $c_{\text {out }}(t)$ and $c_{\text {skip }}(t)$ are differentiable functions such that $c_{\text {out }}(0)=0$, and $c_{\text {skip }}(0)=1$.

We adopt a U-Net architecture enhanced with multi-resolution attention \cite{dhariwal2021diffusion} to serve as the noise estimator network with parameters $\theta$. Conditional guidance is achieved through channel concatenation of \(z_t\) and \(\widetilde{z}\). Additionally, we utilize a VQGAN \cite{esser2021taming} as the encoder \(E\) and decoder \(D\) to convert data between the pixel and latent spaces.

\subsection{Dual-Path Sampling}
We introduce a dual-path architecture for LCM sampling to balance global bone suppression and the retention of local details. As illustrated in Fig. \ref{frame} (b), this architecture comprises a global path (top) implementing global bone suppression and a local path (bottom) focusing on preserving the pulmonary texture details.

\textbf{Local-Enhanced Guidance.} To implement dual-path sampling, we apply different conditions to the conditional LCM. Initially, CXR and masked CXR data were intended to guide the global and local paths, respectively. However, we observed that solely using a local condition in the local path could result in potential boundary artifacts and detail blurring, significantly diminishing the local path’s output quality. This issue may arise from insufficient local guidance, and unnecessary control outside the lung region. Therefore, we recommend enhancing local guidance while reducing global control to address these challenges effectively.

Classifier-Free Guidance (CFG) \cite{ho2021classifier} is a widely used technique in diffusion models to enhance the guidance strength and the quality of generated outputs. This is achieved by incorporating both conditional and unconditional scenarios during the training phase and subsequently adjusting the weights of their predicted outcomes during the inference phase. Drawing inspiration from CFG, we propose LEG in our local-path sampling methodology, which is defined as:

\begin{equation}
\nabla_{z_{l, t}} \log p\left(z_{l, t} \mid \widetilde{z}_l, \widetilde{z}_g\right)= \alpha_{l} \cdot \nabla_{z_{l, t}} \log p\left(z_{l, t} \mid \widetilde{z}_l\right) + (1-\alpha_{l}) \cdot \nabla_{z_{l, t}} \log p\left(z_{l, t} \mid \widetilde{z}_g\right),
\label{eq_leg}
\end{equation}
where \( z_{l, t} \) represents the sample at a given timestep \( t \). The terms \( \widetilde{z}_l \) and \( \widetilde{z}_g \) refer to the local and global conditions, respectively, while \( \alpha_{l} \) denotes the weight associated with the local condition. In our GL-LCM, LEG can be applied without necessitating additional training. This is because the dual-path sampling mechanism inherently encompasses training for both global and local conditions, thereby streamlining the integration of LEG into the model.

\subsection{Global-Local Fusion}
To seamlessly combine the global and local outcomes derived from dual-path sampling, we employ the Poisson Fusion method for global-local integration. This approach effectively preserves the gradient information within the local result, ensuring the retention of fine textures and edges. By utilizing image gradients, Poisson Fusion facilitates a smooth transition between the global low-frequency information and the local high-frequency details. This results in a final output with exceptional global bone suppression performance and remarkable local detail retention. Poisson Fusion is implemented by solving the following equations:

\begin{equation}
\begin{cases} 
\nabla \cdot (\nabla R) = \nabla \cdot (\nabla S_l), & \text{in } M(I_g) \\
R = S_g, & \text{outside of } M(I_g),
\end{cases}
\label{eq_glf}
\end{equation}
where $R$, $S_l$, $S_g$, and $M(I_g)$ represent the fusion result, the local-path result, the global-path result, and the lung region mask calculated by the Dense-U-Net from the CXR image, respectively. The boundary condition is defined as $R = S_g$.

\section{Experiments}

\subsection{Data and Implementation}

\textbf{Data Preparation.}
Acquiring paired CXR and DES soft tissue images is challenging due to the significant scarcity of bone suppression datasets. To address this gap, we utilized the SZCH-X-Rays dataset, a large and high-quality collection from our partner hospital, along with JSRT \cite{shiraishi2000development}, the largest publicly available bone suppression dataset to date, for comprehensive evaluation.

\textbf{The SZCH-X-Rays Dataset} comprises 741 pairs of posterior-anterior CXR and DES soft tissue images, obtained using a dual-exposure DES digital radiography system (Discovery XR656, GE Healthcare) from our partner hospital. Images affected by operational errors, significant motion artifacts, pleural effusion, and pneumothorax were excluded. This dataset is partitioned into 592 training images, 74 validation images, and 75 test images. All images were resized to a resolution of 1024 $\times$ 1024 pixels for consistency.

\textbf{The JSRT Dataset} includes 241 pairs of CXR and soft tissue images sourced from 14 medical centers. These images are categorized into 192 training images, 24 validation images, and 25 test images. Each image was resized to a resolution of 1024 $\times$ 1024 pixels and then transformed into a negative format.

\textbf{Implementation Details.} All experiments were carried out using PyTorch 2.0.1 on a single Nvidia A100 80G GPU within Ubuntu 20.04. The noise estimator network for GL-LCM was trained from scratch over 600 epochs (batch size of 4 and AdamW optimizer). Similarly, the VQGAN model integrated within GL-LCM was trained from scratch for 300 epochs (batch size of 4 and Adam optimizer). A dynamic learning rate schedule was applied in both cases, initializing at 0.0001. The scheduling parameter \(\beta_t\) varied from 0.00085 to 0.012, distributed across \(T = 50\) total sampling timesteps using a scaled linear noise schedule. Finally, the weight $\alpha_l$ in LEG was set to 3.

\subsection{Experimental Results}

\begin{figure}[!tbp]
\centering
\includegraphics[width=0.9\textwidth]{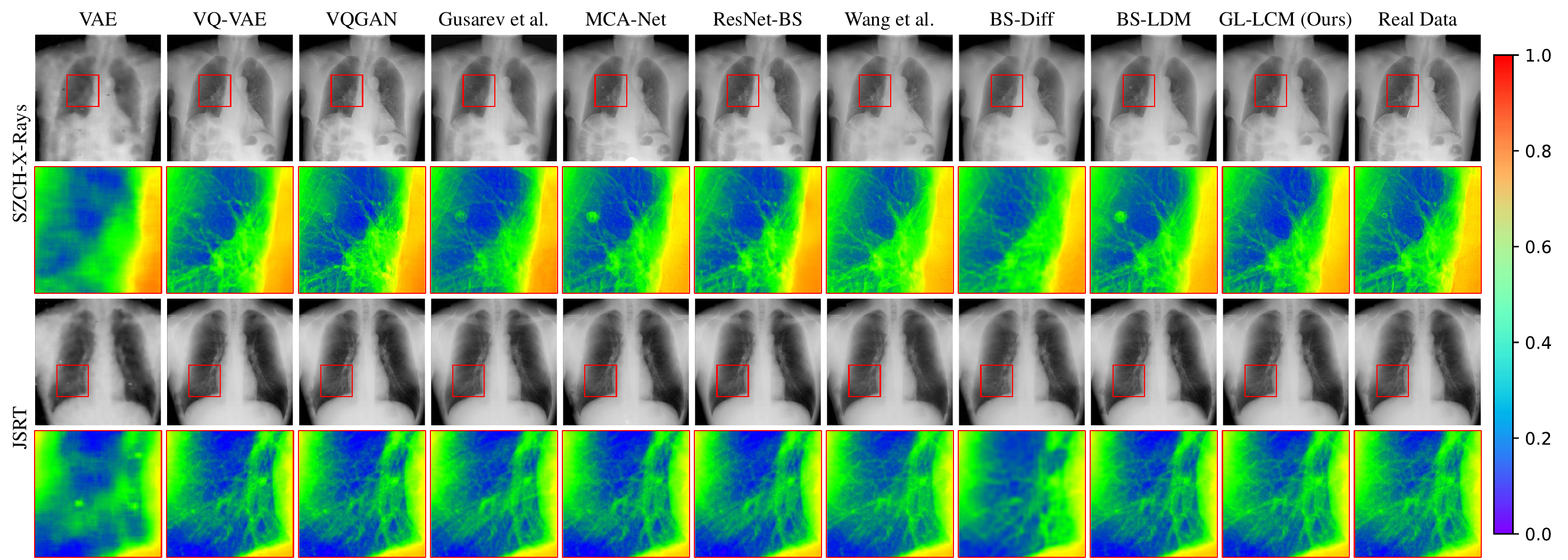}
\caption{Bone suppression and pseudo-color zoomed-in views of GL-LCM compared with state-of-the-art approaches on SZCH-X-Rays (top) and JSRT (bottom) datasets.}
\label{comp}
\end{figure}

\begin{table}[!tbp]
\caption{Comparison of our method with other universal and task-specific approaches in the bone suppression task on the SZCH-X-Rays dataset. The best and second-best results are shown in \textcolor{red}{red} and \textcolor{blue}{blue}, respectively.}\label{tab1}
\centering
\fontsize{8pt}{10pt}\selectfont
\begin{tabular}{>{\arraybackslash}p{2.4cm}>{\centering\arraybackslash}p{2.3cm}>{\centering\arraybackslash}p{2.3cm}>{\centering\arraybackslash}p{2.3cm}>{\centering\arraybackslash}p{2.3cm}}
\toprule
Method  &  BSR (\%)$\uparrow$ & MSE ($10^{-3}$)$\downarrow$ & PSNR$\uparrow$ & LPIPS$\downarrow$\\
\midrule
\multicolumn{5}{l}{Universal Method}\\
\midrule
VAE \cite{kingma2019introduction} &  91.281 $\pm$ 3.088 & 1.169 $\pm$ 1.059 & 30.018 $\pm$ 2.007 & 0.237 $\pm$ 0.047 \\
VQ-VAE \cite{van2017neural} &  94.485 $\pm$ 2.407  & 0.645 $\pm$ 0.596 & 32.600 $\pm$ 2.071 & 0.137 $\pm$ 0.029 \\
VQGAN \cite{esser2021taming} &  94.330 $\pm$ 3.402  & 0.923 $\pm$ 2.478 & 32.096 $\pm$ 2.420 & 0.083 $\pm$ 0.020 \\
\midrule
\multicolumn{5}{l}{Task-Specific Method}\\
\midrule
Gusarev \textit{et al.} \cite{gusarev2017deep} &  94.142 $\pm$ 2.666   & 1.028 $\pm$ 2.201 & 31.369 $\pm$ 2.385 & 0.156 $\pm$ 0.031\\
MCA-Net \cite{zhou2019generation} &  \textcolor{blue}{95.442 $\pm$ 2.095}  & \textcolor{blue}{0.611 $\pm$ 0.435} & \textcolor{blue}{32.689 $\pm$ 1.939} & 0.079 $\pm$ 0.018 \\
ResNet-BS \cite{rajaraman2021chest} &  94.508 $\pm$ 1.733  & 0.646 $\pm$ 0.339 & 32.265 $\pm$ 1.635 & 0.107 $\pm$ 0.022 \\
Wang \textit{et al.} \cite{wang2025rib} &  89.767 $\pm$ 6.079  & 1.080 $\pm$ 0.610 & 29.963 $\pm$ 1.378 & 0.072 $\pm$ 0.016 \\
BS-Diff \cite{chen2024bs} &  92.428 $\pm$ 3.258  & 0.947 $\pm$ 0.510 & 30.627 $\pm$ 1.690 & 0.212 $\pm$ 0.041 \\
BS-LDM \cite{sun2024bs} &  94.159 $\pm$ 2.751  & 0.701 $\pm$ 0.293 & 31.953 $\pm$ 1.969 & \textcolor{blue}{0.070 $\pm$ 0.018} \\
\midrule
GL-LCM (Ours) &  \textcolor{red}{95.611 $\pm$ 1.529}  & \textcolor{red}{0.512 $\pm$ 0.293} & \textcolor{red}{33.347 $\pm$ 1.829} & \textcolor{red}{0.056 $\pm$ 0.015}\\

\bottomrule
\end{tabular}
\end{table}

\begin{table}[!tbp]
\caption{Comparison of our method with other universal and task-specific approaches in the bone suppression task on the JSRT dataset. The best and second-best results are shown in \textcolor{red}{red} and \textcolor{blue}{blue}, respectively.}\label{tab2}
\centering
\fontsize{8pt}{10pt}\selectfont
\begin{tabular}{>{\arraybackslash}p{2.4cm}>{\centering\arraybackslash}p{2.3cm}>{\centering\arraybackslash}p{2.3cm}>{\centering\arraybackslash}p{2.3cm}>{\centering\arraybackslash}p{2.3cm}}
\toprule
Method  &  BSR (\%)$\uparrow$ & MSE ($10^{-3}$)$\downarrow$ & PSNR$\uparrow$ & LPIPS$\downarrow$\\
\midrule
\multicolumn{5}{l}{Universal Method}\\
\midrule
VAE \cite{kingma2019introduction} &  85.646 $\pm$ 9.327  & 1.224 $\pm$ 0.749 & 29.814 $\pm$ 2.364 & 0.155 $\pm$ 0.032 \\
VQ-VAE \cite{van2017neural} &  86.445 $\pm$ 8.881  & 0.986 $\pm$ 0.596 & 30.712 $\pm$ 2.273 & 0.062 $\pm$ 0.017 \\
VQGAN \cite{esser2021taming} &  86.594 $\pm$ 8.916  & 1.002 $\pm$ 0.606 & 30.635 $\pm$ 2.255 & 0.061 $\pm$ 0.017 \\
\midrule
\multicolumn{5}{l}{Task-Specific Method}\\
\midrule
Gusarev \textit{et al.} \cite{gusarev2017deep} & 89.283 $\pm$ 8.288 & 0.821 $\pm$ 0.570 & 31.700 $\pm$ 2.594 & 0.100 $\pm$ 0.024 \\
MCA-Net \cite{zhou2019generation} &  86.887 $\pm$ 9.825  & 0.876 $\pm$ 0.625 & 31.577 $\pm$ 2.905 & \textcolor{blue}{0.057 $\pm$ 0.017} \\
ResNet-BS \cite{rajaraman2021chest} &  88.782 $\pm$ 8.905  & 0.960 $\pm$ 0.661 & 31.021 $\pm$ 2.576 & 0.060 $\pm$ 0.016 \\
Wang \textit{et al.} \cite{wang2025rib} &  \textcolor{blue}{89.679 $\pm$ 9.477}  & 1.013 $\pm$ 0.655 & 30.681 $\pm$ 2.431 & 0.075 $\pm$ 0.015 \\
BS-Diff \cite{chen2024bs} &  88.707 $\pm$ 8.859  & 1.003 $\pm$ 0.655 & 30.765 $\pm$ 2.504 & 0.154 $\pm$ 0.037 \\
BS-LDM \cite{sun2024bs} &  89.322 $\pm$ 9.562  & \textcolor{blue}{0.783 $\pm$ 0.632} & \textcolor{blue}{32.307 $\pm$ 3.231} & 0.058 $\pm$ 0.017 \\
\midrule
GL-LCM (Ours) &  \textcolor{red}{90.056 $\pm$ 10.635}  & \textcolor{red}{0.746 $\pm$ 0.680} & \textcolor{red}{32.951 $\pm$ 3.799} & \textcolor{red}{0.052 $\pm$ 0.015} \\

\bottomrule
\end{tabular}
\end{table}

\begin{table}[!btp]
\caption{Inference efficiency comparison of our method and other diffusion-based methods on the SZCH-X-Rays dataset. The best and second-best results are shown in \textcolor{red}{red} and \textcolor{blue}{blue}, respectively.}\label{tab3}
\centering
\fontsize{8pt}{10pt}\selectfont
\begin{tabular}{>{\arraybackslash}p{2.4cm}>{\centering\arraybackslash}p{2.2cm}>{\centering\arraybackslash}p{2.2cm}>{\centering\arraybackslash}p{2.2cm}>{\centering\arraybackslash}p{2.6cm}}
\toprule
Method  &  Sampler & Sampling Steps & Parameters & Inference Time (s)\\
\midrule
BS-Diff \cite{chen2024bs} &  DDPM  & \textcolor{blue}{1000} & \textcolor{red}{254.7M} & 108.86 \\
BS-LDM \cite{sun2024bs} &  DDPM  & \textcolor{blue}{1000} & \textcolor{blue}{421.3M} & \textcolor{blue}{84.62} \\
GL-LCM (Ours) &  LCM  & \textcolor{red}{50} & 436.9M & \textcolor{red}{8.54} \\
\bottomrule
% \multicolumn{5}{l}{\small $*$ Only the global path of GL-LCM used for inference.}\\
\end{tabular}
\end{table}

\textbf{Comparisons.} We benchmark our method against AE-based methods such as VAE \cite{kingma2019introduction}, VQ-VAE \cite{van2017neural} and Gusarev \textit{et al.} \cite{gusarev2017deep}, GAN-based methods like VQGAN \cite{esser2021taming} and MCA-Net \cite{zhou2019generation}, ResNet-based ResNet-BS \cite{rajaraman2021chest}, and U-Net-based Wang \textit{et al.} \cite{wang2025rib}. We also compare with diffusion-based BS-Diff \cite{chen2024bs}, and BS-LDM \cite{sun2024bs}. For evaluation, we use Bone Suppression Ratio (BSR) \cite{BSR} as task-specific metric, along with three image quality assessment metrics: Mean Square Error (MSE), Peak Signal-to-Noise Ratio (PSNR) \cite{huynh2008scope}, and Learned Perceptual Image Patch Similarity (LPIPS) \cite{zhang2018unreasonable}. 
% BSR, MSE, and PSNR measure pixel-level differences while LPIPS measures perceptual-level differences.

Table \ref{tab1} presents a quantitative comparison of GL-LCM’s performance against these existing methods. Prior AE-based methods, such as VQ-VAE \cite{van2017neural} and Gusarev \textit{et al.} \cite{gusarev2017deep}, show limited success due to their simplistic structures, which struggle with high-frequency details. Although diffusion-based techniques, such as BS-Diff \cite{chen2024bs} and BS-LDM \cite{sun2024bs}, enhance image quality through iterative refinements, they fail to balance bone suppression and detail retention effectively. 

GL-LCM stands out among current state-of-the-art methods. It achieves a 16.20\%, 0.66 dB, and 20.00\% improvement in MSE, PSNR, and LPIPS, respectively, on the SZCH-X-Rays dataset. On the JSRT dataset, it demonstrates improvements of 5.49\%, 0.64 dB, and 8.77\% over other methods. In terms of BSR, GL-LCM achieves 95.61\% on SZCH-X-Rays and 90.06\% on JSRT. Fig. \ref{comp} illustrates the qualitative results for these datasets. Furthermore, the inference time of GL-LCM is about 10\% of what is required by other diffusion-based methods, as outlined in Table \ref{tab3}.

\begin{figure}[!t]
\centering
\includegraphics[width=\textwidth]{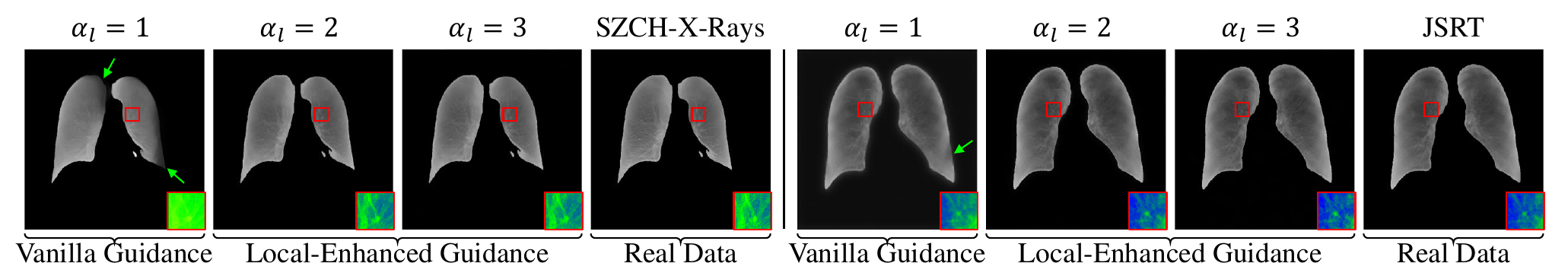}
\caption{The qualitative results of our ablation study of conditional guidance methods of GL-LCM on the SZCH-X-Rays and JSRT datasets. A pseudo-color zoomed-in view is shown in the bottom right corner, and the \textcolor{green}{green} arrows mark the boundary artifacts.}
\label{abl}
\end{figure}

\begin{table}[!t]
\caption{Effect of various conditional guidance methods for local-path sampling on the SZCH-X-Rays and JSRT datasets. The best and second-best results are shown in \textcolor{red}{red} and \textcolor{blue}{blue}, respectively.}\label{tab4}
\centering
\fontsize{8pt}{10pt}\selectfont
\begin{tabular}{>{\centering\arraybackslash}p{3.2cm}>{\centering\arraybackslash}p{2.1cm}>{\centering\arraybackslash}p{2.1cm}>{\centering\arraybackslash}p{2.1cm}>{\centering\arraybackslash}p{2.1cm}}
\toprule
\multirow{2.4}{*}{Guidance Method}
& \multicolumn{2}{c}{SZCH-X-Rays} & \multicolumn{2}{c}{JSRT}\\
\cmidrule(lr){2-3}
\cmidrule(lr){4-5}

& PSNR$\uparrow$ & LPIPS$\downarrow$ & PSNR$\uparrow$ & LPIPS$\downarrow$\\
\midrule
Vanilla Guidance & \textcolor{blue}{32.777 $\pm$ 2.091} & \textcolor{blue}{0.058 $\pm$ 0.016} & 32.296 $\pm$ 3.454 & 0.073 $\pm$ 0.020 \\
CFG \cite{ho2021classifier}& 32.315 $\pm$ 1.717 & 0.068 $\pm$ 0.013 & \textcolor{blue}{32.613 $\pm$ 3.604} & \textcolor{blue}{0.070 $\pm$ 0.015} \\
LEG (Ours) &  \textcolor{red}{33.347 $\pm$ 1.829} & \textcolor{red}{0.056 $\pm$ 0.015} & \textcolor{red}{32.951 $\pm$ 3.799} & \textcolor{red}{0.052 $\pm$ 0.015} \\

\bottomrule
\end{tabular}
\end{table}

\begin{table}[!t]
\caption{Effect of various fusion strategies on the SZCH-X-Rays and JSRT datasets. The best and second-best results are shown in \textcolor{red}{red} and \textcolor{blue}{blue}, respectively.}\label{tab5}
\centering
\fontsize{8pt}{10pt}\selectfont
\begin{tabular}{>{\centering\arraybackslash}p{3.2cm}>{\centering\arraybackslash}p{2.1cm}>{\centering\arraybackslash}p{2.1cm}>{\centering\arraybackslash}p{2.1cm}>{\centering\arraybackslash}p{2.1cm}}
\toprule
\multirow{2.4}{*}{Fusion Strategy}
& \multicolumn{2}{c}{SZCH-X-Rays} & \multicolumn{2}{c}{JSRT}\\
\cmidrule(lr){2-3}
\cmidrule(lr){4-5}

& PSNR$\uparrow$ & LPIPS$\downarrow$ & PSNR$\uparrow$ & LPIPS$\downarrow$\\
\midrule
\XSolidBrush & \textcolor{blue}{31.360 $\pm$ 2.079}  & \textcolor{blue}{0.091 $\pm$ 0.020} & 31.638 $\pm$ 3.078 & 0.074 $\pm$ 0.021 \\
$\alpha$-Fusion \cite{porter1984compositing}&  29.781 $\pm$ 1.522  & 0.181 $\pm$ 0.021 & 31.784 $\pm$ 3.043 & 0.092 $\pm$ 0.013 \\
AE Fusion \cite{yao2019unsupervised} &  30.850 $\pm$ 1.806 & 0.141 $\pm$ 0.028 & \textcolor{blue}{31.835 $\pm$ 3.075} & \textcolor{blue}{0.061 $\pm$ 0.017} \\
Poisson Fusion (Ours) &  \textcolor{red}{33.347 $\pm$ 1.829} & \textcolor{red}{0.056 $\pm$ 0.015} & \textcolor{red}{32.951 $\pm$ 3.799} & \textcolor{red}{0.052 $\pm$ 0.015} \\
\bottomrule
\end{tabular}
\end{table}

\textbf{Ablation Study.} Our ablation study on SZCH-X-Rays and JSRT demonstrates the significant improvements brought by our approach. Table \ref{tab4} and Table \ref{tab5} present the quantitative results.  The incorporation of LEG improves PSNR by 0.57 dB and 0.34 dB and reduces LPIPS by 3.45\% and 25.71\% on SZCH-X-Rays and JSRT, respectively. Unlike Vanilla Guidance solely using a local condition, LEG substantially enhances detail clarity and removes boundary artifacts, as shown in Fig. \ref{abl}. Moreover, our global-local fusion strategy improves PSNR by 1.99 dB and 1.12 dB and reduces LPIPS by 38.46\% and 14.75\% on SZCH-X-Rays and JSRT, respectively. For comparison, the absence of a fusion strategy (LCM baseline) is evaluated using only the global path. By combining the global-local design with LEG, our GL-LCM achieves superior performance overall.

\section{Conclusion}
In this study, we introduce a GL-LCM architecture, which is specifically designed to achieve fast and high-resolution bone suppression while preserving texture details. We also present LEG for effective guidance of local-path sampling, which mitigates potential boundary artifacts and detail blurring. Extensive experiments on self-collected and public datasets demonstrate that GL-LCM delivers impressive performance and efficiency, significantly surpassing many competitive methods. Looking ahead, we plan to further optimize the model by lightweighting its backbone and explore downstream applications, such as computer-aided diagnosis and clinical validation.

\begin{credits}
\subsubsection{\ackname} This work was funded by the National Undergraduate Innovation and Entrepreneurship Training Program of China (No. 202410336081), National Natural Science Foundation of China (No. 61702146, 62076084), and Guangdong Basic and Applied Basic Research Foundation (No. 2025A1515011617, 2022A1515110570). 

\subsubsection{\discintname}
The authors have no competing interests to declare that are relevant to the content of this article.
\end{credits}

\bibliographystyle{splncs04}
\bibliography{main}
\end{document}